\documentclass[aps, prl, preprint]{revtex4-2}

\usepackage{bm, amsmath, amsfonts}
\usepackage[pdftex]{graphicx}

\begin{document}

\title{Pattern Propagation Driven by Surface Curvature}

  \author{Ryosuke Nishide${}^{1,}$}
    \email{r-nishide2018@g.ecc.u-tokyo.ac.jp}

  \author{Shuji Ishihara${}^{1,2,}$}
    \email{csishihara@g.ecc.u-tokyo.ac.jp}

  \affiliation{
  ${}^1$~Graduate School of Arts and Sciences, The University of Tokyo, Komaba 3-8-1, Meguro-ku, Tokyo 153-8902, Japan \\
  ${}^2$~Universal Biology Institute, The University of Tokyo, Komaba 3-8-1, Meguro-ku, Tokyo 153-8902, Japan
  }

\begin{abstract}

  Pattern dynamics on curved surfaces are found everywhere in nature.
  The geometry of surfaces have been shown to influence dynamics and play a functional role, yet a comprehensive understanding is still elusive.
  Here, we report for the first time that a static Turing pattern on a flat surface can propagate on a curved surface,
  as opposed to previous studies,
  where the pattern is presupposed to be static irrespective of the surface geometry.
  To understand such significant changes on curved surfaces,
  we investigate reaction-diffusion systems on axisymmetric curved surfaces.
  Numerical and theoretical analyses reveal that both the symmetries of the surface and pattern participate in the initiation of pattern propagation.
  This study provides a novel and generic mechanism of pattern propagation that is caused by surface curvature, as well as insights into the general role of surface geometry.

\end{abstract}

\maketitle
Pattern formation and dynamics on curved surfaces are ubiquitous,
particularly in biological systems \cite{tan2020topological, Alonso2016, Bischof2017}.
Recent studies reveal the functional roles of the topology and geometry of surfaces in pattern formation~\cite{maroudas2021topological, pieuchot2018curvotaxis, horibe2019curved, saito2021}.
For example, defect dynamics on closed curved surfaces are constrained by the Poincar\'e--Hopf theorem and have been investigated in liquid crystals~\cite{Kralj2011},
flocking~\cite{Shankar2017topological}, and active nematics~\cite{keber2014topology}.
Such defects in cortical actin fibers serve as organization centers in the morphogenesis of hydra regeneration~\cite{maroudas2021topological}.
Molecules such as Bin/Amphiphysin/Rvs domain proteins sense the curvature of a cellular membrane and regulate the cellular shape~\cite{Antonny2011}.
Cellular migration is guided by the curvature of a substrate, which is a response known as ``curvotaxis'' ~\cite{pieuchot2018curvotaxis}.
It was theoretically shown that surface curvature can induce splitting~\cite{horibe2019curved} and rectification~\cite{Davydov2000} of excitable waves.
Rectification by curved surfaces has been reported in the collective motion of self-propelled particles~\cite{ai2019collective}.
However, a comprehensive and general understanding of the effect of surface geometry on pattern dynamics and its functional role remains elusive.

Among pattern formation, Turing patterns are a prominent example arising from reaction-diffusion
systems~\cite{turing1952chemical, kondo2010reaction, raspopovic2014digit, ishihara2006turing}.
Turing patterns on curved surfaces~\cite{Plaza2004, Vandin2016, Charette2019},
such as spheres~\cite{turing1952chemical, varea1999turing, matthews2003pattern, nunez2017diffusion, lacitignola2017turing, sanchez2019turing},
hemispheres~\cite{liaw2001turing, nagata2013reaction},
toruses~\cite{nampoothiri2017role, sanchez2019turing},
ellipsoids~\cite{nampoothiri2017role, nampoothiri2019effect},
and deformed (but axisymmetric) cylinders and spheres~\cite{frank2019pinning} have been investigated previously.
These studies revealed how the Turing instability condition changes
from the flat plane case and how the position of the pattern is modulated by
the inhomogeneity of the surface curvature~\cite{Vandin2016},
which is referred to as ``pinning'' by Frank et al.~\cite{frank2019pinning}
It is noteworthy that in these studies,
the Turing pattern,
which is static on a flat plane,
was assumed to remain static irrespective of the surface geometry.
However, this assumption has not been validated so far~\cite{Krause2021}.

We conducted numerical simulations using the Brusselator~\cite{Prigogine1977} and Lengyel--Epstein (LE) models~\cite{Lengyel1991, Bansagi2015} on several curved surfaces with a parameter set indicating Turing instability for flat planes.
We observed that the static pattern on a flat surface becomes a propagating pattern on several curved surfaces (see Fig.~\ref{'fig1'}),
indicating that a surface curvature can result in the dynamic motion of patterns,
unlike what has been thought thus far.
The simulation results suggest that the surface symmetry contributes to the propagation;
for example, the pattern remains static on a reflection-symmetric cylindrical surface
(see Fig.~\ref{'fig1'}(b)),
whereas a moving pattern appears on a cylindrical surface without reflection symmetry (see Fig.~\ref{'fig1'}(d)).
In addition, propagating waves are observed in both the Brusselator and LE
models, which suggests a generic mechanism underlies this phenomenon.

To systematically investigate the propagation dynamics based on the surface curvature,
we analyzed a reaction-diffusion system on an axisymmetric surface parameterized as ${\bm r} = (x, r(x) \cos \theta, r(x) \sin \theta)$.
$x$ and $\theta$ are defined as $-2\pi \leq x < 2\pi$ and $0 \leq \theta < 2\pi$, respectively,
for which periodic boundary conditions are employed.
Unless otherwise mentioned, we set $r(x) = d+k_1\cos(x)+k_2\cos(2x-\gamma\pi/2)$ with $d = 1.7$, $k_1 = 0.3$, and $k_2 = 0.05$ (see Fig.~\ref{'fig2'}(a)).
$\gamma $ controls the reflection symmetry of the surface about $x=0$; i.e.,
$r(x)=r(-x)$ holds at $\gamma = 0$ but not for $\gamma \neq 0$ (see Fig.~\ref{'fig2'}(b)).
{Herein, we focus on the numerical results obtained using the Brusselator model.
\begin{align}
\label{eq:Brusselator}
  \begin{split}
	\partial_tu &= D_u\Delta u+u^2v-bu-u+a~, \\
    \partial_tv &= D_v\Delta v-u^2v+bu ~,
    \end{split}
\end{align}
where $u$ and $v$ represent chemical concentrations and are functions of the position on the surface $(x,\theta)$ and time $t$.
On a curved surface, the diffusion terms in Equation~(\ref{eq:Brusselator}) are described by the Laplace--Beltrami operator $\Delta$~\cite{Plaza2004, Krause2019}.
For the axisymmetric surface, $\Delta$ is expressed as
\begin{align}
	\label{eq:Laplace-Beltrami}
  \Delta\bullet =
          \frac{1}{r\sqrt{1+r'^2}} \partial_{x}\bigg( \frac{r}{\sqrt{1+r'^2}} \partial_{x}\bullet\bigg)
          +\frac{1}{r^2} \partial^2_{\theta}\bullet~,
\end{align}
where $r'$ represents $dr(x)/dx$.
The surface on a normal cylinder is equivalent to a flat plane, since they have the same metric.
By choosing the appropriate parameter set,
the system becomes Turing-unstable,
as indicated by the dispersion relation $\mu(\lambda)$ and numerical simulations on a flat plane (see Figs.~\ref{'fig2'}(c) and (d)).
The dispersion relation $\mu(\lambda)$ is obtained via linear stability analysis in the uniform state $(u,v) = (a,b/a)$
and represents the growth rate of a mode characterized by $\lambda$, where $\lambda$ is an eigenvalue of the
Laplace--Beltrami operator determined by
$\Delta \phi = -\lambda \phi$.
For flat surfaces, $\lambda$ coincides with the square of the wavenumber.
$\mu(\lambda)$ in Fig.~\ref{'fig2'}(c) takes real positive values between a finite range of $\lambda$,
exhibiting a typical form of Turing instability.
Note that the dispersion relation $\mu(\lambda)$,
and thus the Turing condition, is the same for any surface~\cite{Plaza2004}.

We performed numerical simulations by varying $\gamma$.
When the pattern is helical stripes on a reflection-symmetric surface,
the dynamics become static,
as shown in Fig.~\ref{'fig3'}(a) ($\gamma=0$).
As the reflection symmetry of the surface is diminished owing to changes in $\gamma$,
pattern propagation emerges, as shown in Fig.~\ref{'fig3'}(a) ($\gamma = 1$).
The velocity of the propagation is plotted against $\gamma$ in Fig.~\ref{'fig3'}(b) (blue points).
The propagation velocity is proportionally dependent on $\gamma$ in the vicinity of $\gamma=0$,
indicating that slight reflection asymmetry is sufficient to trigger propagation.
Depending on the parameter and initial conditions,
stripes almost parallel along the $x$-axis can appear (see Fig.~\ref{'fig3'}(c)).
For such a pattern, we do not observe propagation for any value of $\gamma$.

Dotted patterns also appeared via Turing instability.
On the axisymmetric surface, we observed dotted patterns that are aligned helically (Fig.~\ref{'fig3'}(d))
and parallelly along the $x$-axis (Fig.~\ref{'fig3'}(f)).
The helically aligned dots pattern remains static on the reflection-symmetric surface at $\gamma = 0$
but propagates at $\gamma \neq 0$ (see Figs.~\ref{'fig3'}(d) and (e)).
By contrast, the parallelly aligned dots patterns remain
static for any value of $\gamma$ (see Fig.~\ref{'fig3'}(f)).
Similar results for stripe and dotted patterns are obtained in simulations based on other parameter sets
and in the LE model,
as well as on surfaces with different forms of $r(x)$ (see Figs.~\ref{'fig1'}(b) and (d)).
These observations suggest that the reflection asymmetry of the surface and
profile of the pattern are both responsible for the initiation of pattern propagation.

To analyze these numerical results, we examined the observed patterns in terms of symmetry;
see the projected concentration profile of $u(x,\theta)$ at a late time point $t$
on the $x$-$\theta$ plane in Figs.~\ref{'fig4'}(a)--(d).
The patterns obtained in the simulations can be classified into two types.
One is referred to as reflection-symmetric patterns about the $\theta$-axis,
where the patterns are exactly invariant by reflection about the $\theta$-axis with an appropriate
reflection axis and translation along the $x$-axis by an integer multiple of $2\pi$,
satisfying $u(x,\theta) = u(x+2n \pi,-\theta+\theta_0)$
with appropriate values of $\theta_0$ and integer $n$ (see Figs.~\ref{'fig4'}(a) and (c)).
Such patterns include the ``parallel stripes'' and ``parallelly aligned dots'' along the $x$-axis
mentioned above (see Fig.~\ref{'fig3'} (c) and (f)), and they remain static.
The other patterns include the ``helical stripes'' and ``helically aligned dots'' (see Fig.~\ref{'fig3'}(a) and (d)) that appear to be inversion-symmetric about the $\theta$- and $x$-axes (see Figs.~\ref{'fig4'}(b) and (d)).
However, the inversion symmetry holds exactly only at $\gamma = 0$, where the patterns satisfy $u(x,\theta) = u(-x+2n\pi,-\theta+\theta_0)$ with appropriate $\theta_0$ and $n$ values.
For $\gamma \neq 0$, the inversion symmetry is merely an approximation.
Thus, the absence of the inversion-symmetric solution is the key to the emergence of propagation.

Based on the feature of symmetry described above,
we performed a theoretical analysis to clarify the mechanism of pattern propagation on curved surfaces.
A reaction-diffusion system is generally expressed as
\begin{align}
	\label{eq:general_RD}
	 \partial_t\bm{U}&=D\Delta \bm{U}+\bm{R}(\bm{U}),
\end{align}
where ${\bm U}$ is a vector composed of the chemical concentration,
$D$ is a diagonal diffusion matrix, and $\bm{R}(\bm U)$ is a vector of reaction terms.
Using variable $\rho \equiv \theta - \omega t$, the propagating solution along the $\theta$-axis
is expressed as ${\bm U}(x,\theta,t) = {\bm U(x,\rho)}$.
Subsequently, Equation~(\ref{eq:general_RD}) for the propagating solution reads
\begin{align}
	\label{eq:general_RD_TWs}
    \omega\partial_\rho\bm{U}+D\Delta \bm{U}+\bm{R}(\bm{U})=0.
\end{align}
By taking the inner product with $\partial_\rho\bm{U}$ and integrating over the surface, we obtain
\begin{align}\label{eq:TWs_velocity}
   \omega = -\frac{\int dS~\partial_\rho\bm{U}^T\bm{R}}{\int dS~\big(\partial_\rho\bm{U}\big)^2}~,
\end{align}
where $dS \equiv r\sqrt{1+r'^2}dxd\rho$ is a surface area element.
This relationship is consistent with the simulation data, as shown by the orange lines in Figs.~\ref{'fig3'}(b) and (e).
Note that for relaxation systems where the reaction terms are expressed by the gradient of an energy function $H(\bm U)$ as ${\bm R}(\bm U) = - \partial H(\bm U)/\partial {\bm U}$,
one can easily prove that $\omega = 0$ using Equation~(\ref{eq:TWs_velocity}); this indicates that propagation is realizable only in out-of-equilibrium systems.

Equation~(\ref{eq:TWs_velocity}) relates propagation velocity to the pattern profile and surface geometry.
Considering a pattern with reflection symmetry about the $\theta$-axis (a pattern satisfying
${\bm U}(x,\rho) = {\bm U}(x+2n\pi,-\rho+\rho_0)$ with appropriate $n$ and $\rho_0$),
$\bm{R}(\bm U)$ satisfies the same symmetry.
Subsequently, owing to the parity of $\partial_{\rho} \bm{U}$ and ${\bm R}(\bm U)$,
the integral in the numerator of the equation vanishes,
and accordingly, $\omega = 0$ for such a pattern.
This applies similarly to the case for a helical inversion-symmetric pattern at $\gamma = 0$,
which satisfies ${\bm U}(x,\rho) = {\bm U}(-x+2n\pi,-\rho+\rho_0)$.
These arguments prove the absence of propagation in
parallel stripes and parallelly aligned dots patterns along the $x$-axis
for any value of $\gamma$, as well as in helical patterns at $\gamma = 0$.
By contrast, for a helical pattern with $\gamma \neq 0$,
these symmetries do not hold, and in general, $\omega$ is finite, as shown below.

To investigate the mechanism by which propagation occurs due to the breaking of reflection symmetry of the surface geometry,
we performed a perturbation analysis by setting the radius of the axisymmetric surface to $r(x) = r_0(x)+\epsilon r_1(x)$.
An even function $r_0(x)$ represents the reflection-symmetric part of the surface, whereas
an odd function $r_1(x)$ represents the asymmetric part.
For a small $\epsilon$, the Laplace--Beltrami operator is expanded to $\Delta=\Delta_0+\epsilon\bar{\Delta}_1+\mathcal{O}(\epsilon^2)$,
where $\Delta_0$ is the operator for reflection-symmetric surfaces, and $\epsilon\bar{\Delta}_1$
represents modulation via asymmetric deformation.
$\Delta_0$ and $\bar{\Delta}_1$ exhibit the opposite parity for reflection about the $x$-axis, i.e.,
they change as $\Delta_0 \to \Delta_0$ and $\bar{\Delta}_1 \to -\bar{\Delta}_1$,
respectively, for a transformation $x \to -x$.
At $\epsilon = 0$, we assume that the system
shows a helical pattern ${\bm U}_0$ satisfying inversion symmetry
${\bm U}_0(x,\rho) = {\bm U}_0(-x+2n\pi,-\rho+\rho_0)$,
for which the velocity vanishes as discussed above.
The effect of asymmetric surface deformation on the pattern propagation dynamics is evaluated as follows:
\begin{align}
	\label{eq:Omega_1}
 \omega = -\epsilon \frac{\int dS_0~\bm{W}_0^TD\bar{\Delta}_1\bm{U}_0}{\int dS_0~\bm{W}_0^T\partial_\rho\bm{U}_0} + \mathcal{O}(\epsilon^2)~.
\end{align}
where $dS_0 \equiv r_0\sqrt{1+r'^2_0}dxd\rho$ and function vector ${\bm W}_0$ are determined by the system at $\epsilon = 0$.
Owing to the parity of $\bar{\Delta}_1$, Equation~(\ref{eq:Omega_1}) does not vanish in general,
indicating that the loss of reflection-symmetry of the surface about the $x$-axis causes the propagation of the Turing pattern.
We numerically verified the relationship above based on our simulation data
by setting $r_0(x)=d+k_1\cos(x)+k_2\cos(2x)$, $r_1(x)=(\pi k_2/2)\sin(2x)$,
and $\epsilon=\gamma$, and discovered good agreement in the vicinity of $\epsilon = 0$ (Figs.~\ref{'fig3'}(b) and (e), dashed green lines).
Taken together, these analyses corroborate the emergence of Turing patterns propagating on curved surfaces.

It is noteworthy that the analyses above do not exclude a moving Turing pattern on a highly symmetric
surface if the pattern is out of symmetry.
In our numerical simulations, we observed a stripe pattern propagating on a spherical surface in some cases, where the pattern does not satisfy the symmetry expected from the surface.

In summary, we discovered chemical waves that propagated genuinely driven by surface curvature.
This propagation does not occur in one-dimensional systems, where no intrinsic curvature exists, which contrasts with typical propagating waves such as those in excitable media, where the initiation of wave propagation is independent of the surface geometry.
By performing numerical simulations and perturbative analysis, we identified the generic conditions for pattern propagation irrespective of the model equations for axisymmetric surfaces,
where loss of surface reflection symmetry along the $x$-axis results in loss of pattern inversion symmetry and propagation along the $\theta$-axis.
The (a)symmetry of the surface and pattern are both important, suggesting that in general surfaces, pattern dynamics is determined by the geometric feature of the surface and pattern profile.
The pattern propagation discovered in this study was overlooked previously,
likely because most of those previous studies focused primarily on highly symmetric surfaces.
In addition, the propagation velocity is generally much slower
than that of pattern formation at the early stage
(see early stages $t<10^3$ of the simulations shown in kymographs of Figs.~\ref{'fig3'}(a) and (d)).

Our findings imply the new roles of surface geometry for pattern dynamics,
applicable to natural and engineering systems.
For example, geometry-dependent information transduction is possible in a growing organ, where
deformation of the surface can cause initiation (or suppression) of wave propagation,
which can subsequently result in the feedback regulation of organ growth.
Similar regulation between pattern and surface geometry is possible in the molecular localization on the cell membrane, either inside or outside of the surface.
In the future, our study should be extended to general curved surfaces and network systems~\cite{Nakao2010},
including deformable surfaces~\cite{Krause2019, Miller2018, Tamemoto2020}.
Furthermore, it would be interesting to investigate similar phenomena in systems other than reaction-diffusion systems, such as active matter systems with polar and nematic orders.

We thank T. Namba, M. Tateno and N. Saito for their fruitful comments.
This study was supported by JSPS KAKENHI (JP18057992), JPJSJPR 20191501,
and JST CREST JPMJCR1923, Japan (to S.I.) and by JST SPRING, grant number JPMJSP2108 (to R.N.).
Both R.N. and S.I. proposed the research direction, contributed to the theoretical analysis, and wrote the manuscript. R.N. performed all numerical simulations.

\begin{figure}[t]
\includegraphics[keepaspectratio,scale=1.0]{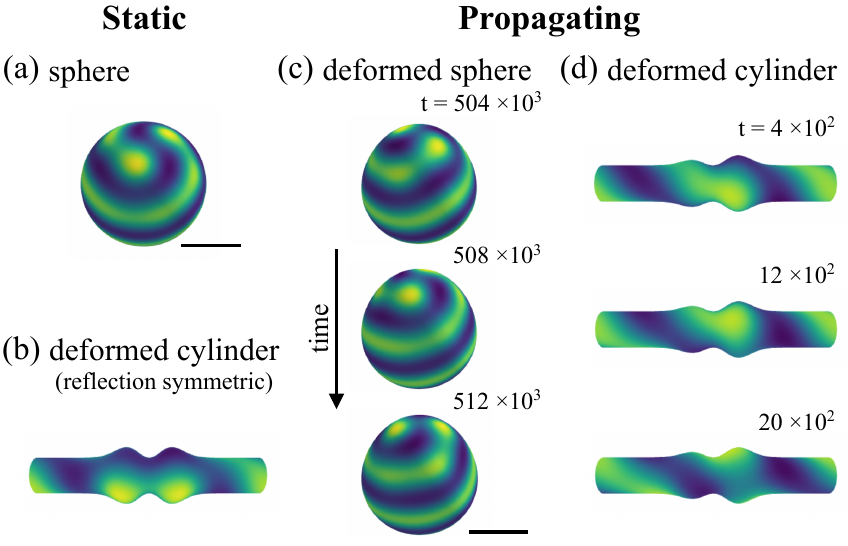}
\caption{
Propagating and static patterns on surfaces.
Concentration $u$ obtained from numerical simulations of Brusselator model are indicated by color scale.
Lighter (darker) colors represent higher (lower) concentrations.
Parameters are set as $(a,b,D_u,D_v)=(2.0,4.5,0.5,1.8)$.
(a,b) Static patterns on sphere (a) and reflection-symmetric deformed cylinder (b).
(c,d) Propagating patterns appear on deformed spheres (c) and deformed cylinders (d).
Periodic boundary conditions are used for cylindrical surfaces.
Scale bars: Six simulation length units (slu).
For deformed cylinders, axial length is $4\pi$.
}
\label{'fig1'}
\end{figure}

\begin{figure}[t]
\includegraphics[keepaspectratio,scale=1.0]{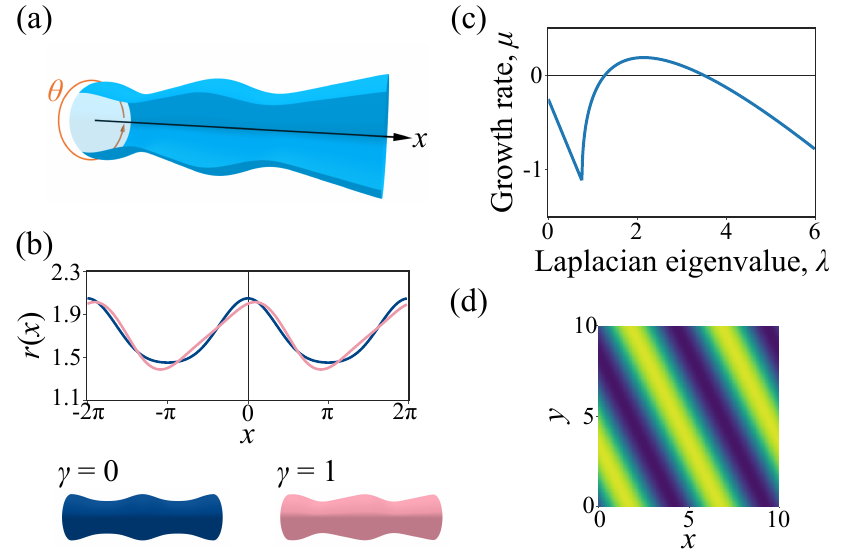}
\caption{
Propagation of stripe patterns on axisymmetric surfaces.
(a,b)~Axisymmetric surface whose radius is $r(x) = d+k_1\cos(x)+k_2\cos(2x-\gamma\pi/2)$.
$\gamma$ represents parameter for controlling surface reflection asymmetry along $x$-axis.
Surfaces at $\gamma=0$ and $1$ are shown by dark blue and light red lines in panel (b), respectively.
(c)~Dispersion relation with parameter set $(a,b,D_u,D_v) = (2.0,4.5,0.5,1.8)$.
(d)~Turing pattern on flat plane with parameter set shown in (c).
}
\label{'fig2'}
\end{figure}

\begin{figure*}[t]
\includegraphics[keepaspectratio,scale=1.04]{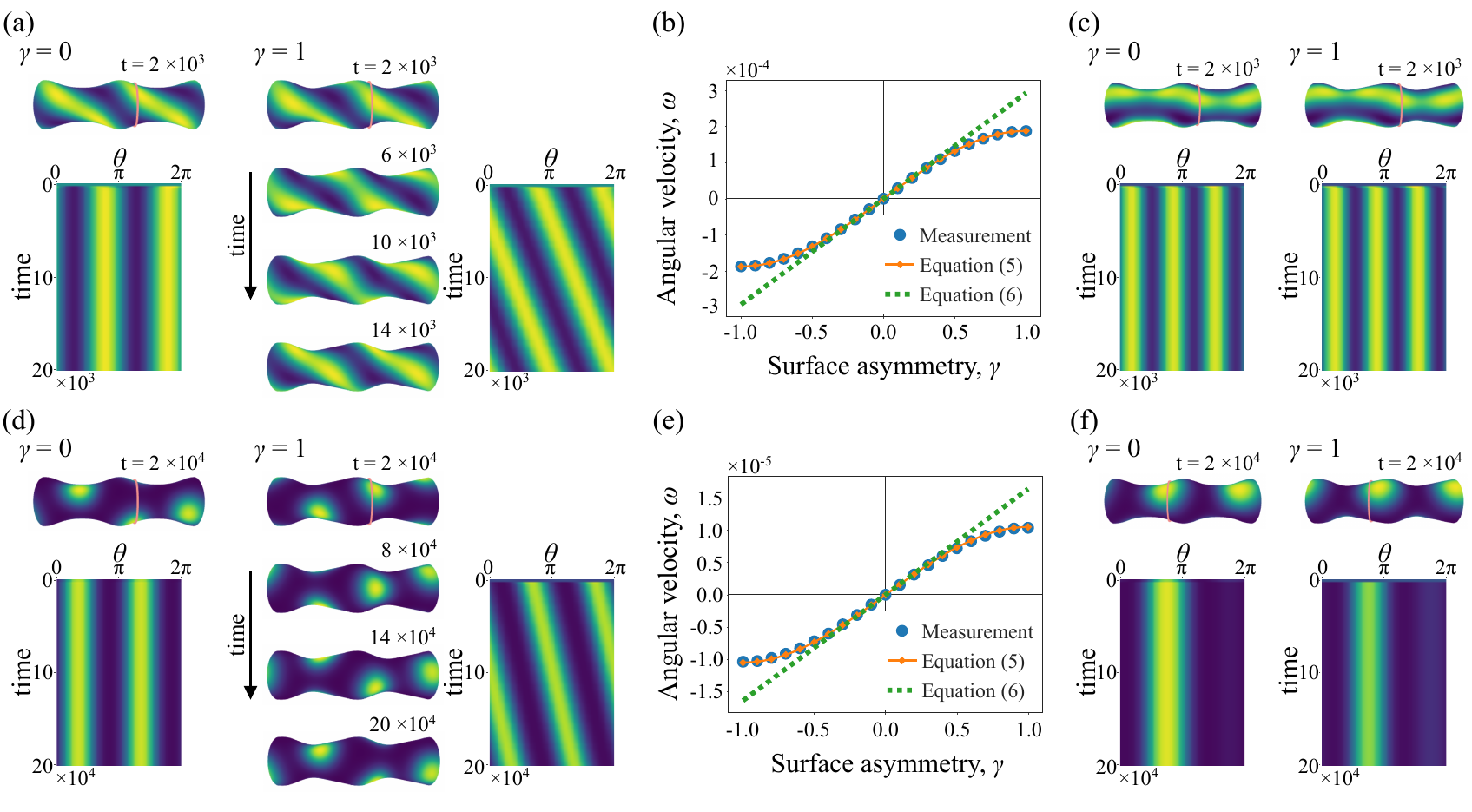}
\caption{
Propagation of stripe and dotted patterns on axisymmetric surfaces.
(a)~Helical stripe patterns on reflection symmetric ($\gamma=0$) and asymmetric ($\gamma = 1$) surfaces.
Kymographs of these patterns along $\theta$-axis (indicated by pale red line) are shown.
(b)~Angular velocity $\omega$ of propagating pattern along $\theta$-axis against surface asymmetry $\gamma$.
$\omega$ is obtained via three different methods:~direct measurement (blue point),
velocity relation equation~(\ref{eq:TWs_velocity}) (orange line), and perturbative velocity
relation equation~(\ref{eq:Omega_1}) (dashed green line).
(c)~Parallel stripes patterns along $x$-axis on reflection symmetric (left) and asymmetric (right) surfaces.
(d)--(f) Date equivalent to (a)--(c) for helically aligned dots patterns
((d),(e)) and for parallelly aligned dots patterns (f).
Parameter sets of patterns are $(a,b,D_u,D_v) = (2.0,4.5,0.5,1.8)$ (a),$~(2.8, 5.0, 0.4, 2.4)$ (c),$~(1.5,2.5,0.3,3.0)$ (d), and $(1.5,3.0,0.5,3.5)$ (f).
}
\label{'fig3'}
\end{figure*}

\begin{figure}[t]
\includegraphics[keepaspectratio,scale=1.0]{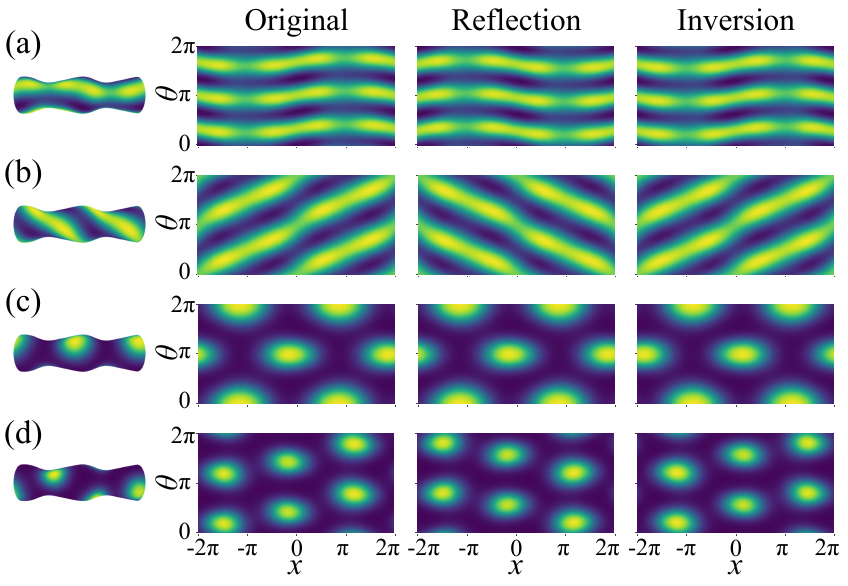}
\caption{
Relationship between pattern symmetry and pattern propagation.
(a)--(d),~(first and second columns) patterns on surface and projected images on $x$-$\theta$ planes, respectively.
$\gamma$ is set to $1.0$.
(third column)~patterns obtained by reflecting original patterns along $\theta$-axis.
(fourth column)~patterns obtained by inverting original patterns.
For ease of comparison,
the images in the third and fourth columns are translated along the $x$- and $\theta$-axes
such that the surface is invariant (i.e., $x \to x+2n\pi$ and $\theta \to \theta+ \theta_0$).
(a) parallel stripes along $x$-axis;
(b) helical stripes;
(c) parallelly aligned dots along $x$-axis;
(d) helically aligned dots.
}
\label{'fig4'}
\end{figure}


\begin{thebibliography}{99}

  \bibitem{tan2020topological}
  T. H. Tan, J. Liu, P. W. Miller, M. Tekant, J. Dunkel, and N. Fakhri,
  Topological turbulence in the membrane of a living cell,
  {Nat. Phys.} $\bm{16}$, 657--662 (2020).

  \bibitem{Bischof2017}
  J. Bischof, C. A. Brand, K. Somogyi, I. M\'ajer, S. Thome, M. Mori,
  U. S. Schwarz, and P. L\'en\'art,
  A cdk1 gradient guides surface contraction waves in oocytes,
  {Nat. Commun.} $\bm{8}$, 849 (2017).

  \bibitem{Alonso2016}
  S. Alonso, M. B\"ar, and B. Echebarria,
  Nonlinear physics of electrical wave propagation in the heart:~a review,
  {Rep. Prog. Phys.} $\bm{79}$, 096601 (2016).

  \bibitem{maroudas2021topological}
  Y. Maroudas-Sacks, L. Garion, L. Shani-Zerbib, A. Livshits, E. Braun, and K. Keren,
  Topological defects in the nematic order of actin fibres as organization centres of Hydra morphogenesis,
  {Nat. Phys.} $\bm{17}$, 251--259 (2021).

  \bibitem{pieuchot2018curvotaxis}
  L. Pieuchot \textit{et al.},
  Curvotaxis directs cell migration through cell-scale curvature landscapes,
  {Nat. Commun.} $\bm{9}$, 3995 (2018).

  \bibitem{horibe2019curved}
  K. Horibe, K. Hironaka, K. Matsushita, and K. Fujimoto,
  Curved surface geometry-induced topological change of an excitable planar wavefront,
  {Chaos} $\bm{29}$, 093120 (2019).

  \bibitem{saito2021}
  N. Saito, and S. Sawai,
  Three-dimensional morphodynamic simulations of macropinocytic cups,
  {iScience} $\bm{24}$, 103087 (2021).

  \bibitem{Kralj2011}
  S. Kralj, R. Rosso, and E. G. Virga,
  Curvature control of valence on nematic shells,
  {Soft Matter} $\bm{7}$, 670--683 (2011).

  \bibitem{Shankar2017topological}
  S. Shankar, M. J. Bowick, and M. C. Marchetti,
  Topological sound and flocking on curved surfaces,
  {Phys. Rev. X} $\bm{7}$, 031039 (2017).

  \bibitem{keber2014topology}
  F. C. Keber, E. Loiseau, T. Sanchez, S. J. Decamp, L. Giomi,
  M. J. Bowick, M. C. Marchetti, Z. Dogic, and A. R. Bausch,
  Topology and dynamics of active nematic vesicles,
  {Science} $\bm{345}$, 1135--1139 (2014).

  \bibitem{Antonny2011}
  B. Antonny,
  Mechanisms of membrane curvature sensing,
  {Annu. Rev. Biochem.} $\bm{80}$, 101--123 (2011).

  \bibitem{Davydov2000}
  V. A. Davydov, V. G. Morozov, and N. V. Davydov,
  Ring-shaped autowaves on curved surfaces,
  {Phys. Lett. A} $\bm{267}$, 326--330 (2000).

  \bibitem{ai2019collective}
  B. Q. Ai, W. J. Zhu, and J. J. Liao,
  Collective transport of polar active particles on the surface of a corrugated tube,
  {New J. Phys.} $\bm{21}$, 093041 (2019).

  \bibitem{turing1952chemical}
  A. M. Turing,
  The chemical basis of morphogenesis,
  {Phil. Trans. R. Soc. Lond. B} $\bm{237}$, 37--72 (1952).

  \bibitem{kondo2010reaction}
  S. Kondo, and T. Miura,
  Reaction-diffusion model as a framework for understanding biological pattern formation,
  {Science} $\bm{329}$, 1616--1620 (2010).

  \bibitem{raspopovic2014digit}
  J. Raspopovic, L. Marcon, L. Russo, and J. Sharpe,
  Digit patterning is controlled by a Bmp-Sox9-Wnt Turing network modulated by morphogen gradients,
  {Science} $\bm{345}$, 566--570 (2014).

  \bibitem{ishihara2006turing}
  S. Ishihara, and K. Kaneko,
  Turing pattern with proportion preservation,
  {J. Theor. Biol.} $\bm{238}$, 683--693 (2006).

  \bibitem{Plaza2004}
  R. G. Plaza, F. S\'anchez-Gardu\~no, P. Padilla, R. A. Barrio, and P. K. Maini,
  The Effect of Growth and Curvature on Pattern Formation,
  {J. Dyn. Differ. Equ.} $\bm{16}$, 1093--1121 (2004).

  \bibitem{Vandin2016}
  G. Vandin, D. Marenduzzo, A. B. Goryachev, and E. Orlandini,
  Curvature-driven positioning of Turing patterns in phase-separating curved membranes,
  {Soft Matter} $\bm{12}$, 3888--3896 (2016).

  \bibitem{Charette2019}
  L. Charette,
  Pattern formation on curved surfaces,
  Ph.D. Thesis, The University of British Columbia, Canada, (2019).

  \bibitem{varea1999turing}
  C. Varea, J. L. Aragon, and R. A. Barrio,
  Turing patterns on a sphere,
  {Phys. Rev. E} $\bm{60}$, 4588 (1999).

  \bibitem{matthews2003pattern}
  P. C. Matthews,
  Pattern formation on a sphere,
  {Phys. Rev. E} $\bm{67}$, 036206 (2003).

  \bibitem{nunez2017diffusion}
  M. N{\'u}{\~n}ez-L{\'o}pez, G. Chac{\'o}n-Acosta, and J. A. Santiago,
  Diffusion-driven instability on a curved surface: spherical case revisited,
  {Braz. J. Phys.} $\bm{47}$, 231--238 (2017).

  \bibitem{lacitignola2017turing}
  D. Lacitignola, B. Bozzini, M. Frittelli, and I. Sgura,
  Turing pattern formation on the sphere for a morphochemical reaction-diffusion model for electrodeposition,
  {Commun. Nonlinear Sci. Numer. Simul.} $\bm{48}$, 484-508 (2017).

  \bibitem{sanchez2019turing}
  F. S{\'a}nchez-Garduno, A. L. Krause, J. A. Castillo, and P. Padilla,
  Turing-Hopf patterns on growing domains: the torus and the sphere,
  {J. Theor. Biol.} $\bm{481}$, 136--150 (2019).

  \bibitem{liaw2001turing}
  S. S. Liaw, C. C. Yang, R. T. Liu, and J. T. Hong
  Turing model for the patterns of lady beetles,
  {Phys. Rev. E} $\bm{64}$, 041909 (2001).

  \bibitem{nagata2013reaction}
  W. Nagata, H. R. Z. Zangeneh, and D. M. Holloway,
  Reaction-diffusion patterns in plant tip morphogenesis: bifurcations on spherical caps,
  {Bull. Math. Biol.} $\bm{75}$, 2346--2371 (2013).

  \bibitem{nampoothiri2017role}
  S. Nampoothiri, and A. Medhi,
  Role of curvature and domain shape on Turing patterns,
  arXiv:1705.02119 (2017).

  \bibitem{nampoothiri2019effect}
  S. Nampoothiri,
  Effect of geometry on the positioning of a single spot in reaction-diffusion systems,
  arXiv:1909.06528 (2019).

  \bibitem{frank2019pinning}
  J. R. Frank, J. Guven, M. Kardar, and H. Shackleton,
  Pinning of diffusional patterns by non-uniform curvature,
  {EPL} $\bm{127}$, 48001 (2019).

  \bibitem{Krause2021}
  A. L. Krause, E. A. Gaffney, P. K. Maini, and V. Klika,
  Modern perspectives on near-equilibrium analysis of Turing systems,
  {Phil. Trans. R. Soc. A.} $\bm{379}$, 20200268 (2021).

  \bibitem{Prigogine1977}
  G. Nicolis, and I. Prigogine,
  {Self-Organization in Nonequilibrium Systems}
  (Wiley, New York, 1977).

  \bibitem{Lengyel1991}
  I. Lengyel, and I. R. Epstein,
  Modeling of Turing structures in the chlorite-iodide-malonic acid-starch reaction system,
  {Science} $\bm{251}$, 650--652 (1991).

  \bibitem{Bansagi2015}
  T. B\'ansa\'gi Jr., and A. F. Taylor,
  Helical Turing patterns in the Lengyel-Epstein model in thin cylindrical layers,
  {Chaos} $\bm{25}$, 064308 (2015).

  \bibitem{Krause2019}
  A. L. Krause, M. A. Ellis, and R. A. Van Gorder,
  Influence of curvature, growth, and anisotropy on the evolution of Turing patterns on growing manifolds,
  {Bull. Math. Biol.} $\bm{81}$, 759--799 (2019).

  \bibitem{Nakao2010}
  H. Nakao, and A. S. Mikhailov,
  Turing patterns in network-organized activator-inhibitor systems,
  {Nat. Phys.} $\bm{6}$, 544--550 (2010).

  \bibitem{Miller2018}
  P. W. Miller, N. Stoop, and J. Dunkel,
  Geometry of wave propagation on active deformable surfaces,
  {Phys. Rev. Lett.} $\bm{120}$, 268001 (2018).

  \bibitem{Tamemoto2020}
  N. Tamemoto, and H. Noguchi,
  Pattern formation in reaction-diffusion system on membrane with mechanochemical feedback,
  {Sci. Rep.} $\bm{10}$, 19582 (2020).

\end{thebibliography}
\end{document}